\documentclass[sigconf]{acmart}

\usepackage{booktabs} 

\usepackage{amsmath}
\usepackage{amssymb}
\usepackage{listings} 
\usepackage{url}
\usepackage{graphicx}
\usepackage{nameref}
\usepackage{multirow}
\usepackage[inline]{enumitem}
\usepackage{array}
\usepackage{multicol}
\usepackage{graphicx}
\usepackage{diagbox}
\usepackage{color}
\usepackage{listings}
\usepackage{wrapfig}
\usepackage{lscape}
\usepackage{rotating}
\usepackage{epstopdf}
\usepackage{mathtools}
\usepackage{subfig}
\usepackage{makecell}
\usepackage[bottom]{footmisc}

\renewcommand\footnotetextcopyrightpermission[1]{}
\begin{document}

\sloppy

\copyrightyear{2017} 
\acmYear{2017} 
\setcopyright{acmlicensed}
\acmConference{PROMISE }{November 8, 2017}{Toronto, Canada}
\acmPrice{15.00}
\acmDOI{10.1145/3127005.3127016}
\acmISBN{978-1-4503-5305-2/17/11}

\title{Boosting Automatic Commit Classification Into Maintenance Activities By Utilizing Source Code Changes}

\author{Stanislav Levin}
\affiliation{%
\institution{Tel Aviv University}
\department{The Blavatnik School of Computer Science}
\city{Tel Aviv}
\country{Israel}
}
\email{stanisl@post.tau.ac.il}
           
\author{Amiram Yehudai}
\affiliation{%
\institution{Tel Aviv University}
\department{The Blavatnik School of Computer Science}
\city{Tel Aviv}
\country{Israel}
}
\email{amiramy@tau.ac.il}

\begin{abstract}

\noindent \textbf{Background}: Understanding maintenance activities performed in a source code repository could help practitioners reduce uncertainty and improve cost-effectiveness by planning ahead and pre-allocating resources towards source code maintenance. The research community uses 3 main classification categories for maintenance activities: Corrective: fault fixing; Perfective: system improvements; Adaptive: new feature introduction.
Previous work in this area has mostly concentrated on evaluating commit classification (into maintenance activities) models in the scope of a single software project.\\
\textbf{Aims:} In this work we seek to design a commit classification model capable of providing high accuracy and Kappa across different projects. In addition, we wish to compare the accuracy and kappa characteristics of classification models that utilize word frequency analysis, source code changes, and combination thereof. \\
\textbf{Method:} We suggest a novel method for automatically classifying commits into maintenance activities by utilizing source code changes (e.g, statement added, method removed, etc.).
The results we report are based on studying 11 popular open source projects from various professional domains from which we had manually classified 1151 commits, over 100 from each of the studied projects. Our models were trained using 85\% of the dataset, while the remaining 15\% were used as a test set.\\
\textbf{Results:} Our method shows a promising accuracy of 76\% and Cohen's kappa of 63\% (considered ''Good`` in this context) for the test dataset, an improvement of  over 20 percentage points, and a relative boost of $\sim$40\% in the context of cross-project classification.\\
\textbf{Conclusions:} We show that by using source code changes in combination with commit message word frequency analysis we are able to considerably boost classification quality in a project agnostic manner. 
\end{abstract}



\maketitle

\section{Introduction}\label{sec:intro}

Three main classification categories for maintenance activities in software projects were identified by Mockus et al.\cite{mockus2000identifying}:

\begin{itemize}
    \item Corrective: fixing faults, functional and non-functional.
    \item Perfective: improving the system and its design.
    \item Adaptive: introducing new features into the system.
\end{itemize}

Understanding these maintenance activities, performed in a source code repository, could help practitioners reduce uncertainty and improve cost-effectiveness \cite{swanson1976dimensions} by planning ahead and pre-allocating resources towards source code maintenance. 
Maintenance activity profiles of software projects have therefore been a subject of research in numerous works \cite{swanson1976dimensions,mockus2000identifying, meyers1988, lientz1978characteristics, levinIcsme2016, schach2003determining}. 
To determine maintenance activity profiles, one must first classify the activities, which come in the form of developer commits to the version control system (VCS). 
A widely practiced method for commit classification has been inspecting the commit's textual comment field (a.k.a commit message) \cite{mockus2000identifying, fischer2003populating, sliwerski2005changes, amor2006discriminating}. Works employing comment based classification models reported the accuracy to average below 60\% when used in the scope of a single project, and below 53\% when used in the scope of multiple projects (i.e., when a single model was used to classify commits from multiple projects) \cite{hindle2009automatic, amor2006discriminating}.
Our work is motivated by the following observations:

\begin{enumerate}    
    \item Existing results rarely consider cross-project classification, which threatens external validity. Hindle et al. \cite{hindle2009automatic} explored cross-project classification and reported the accuracy to be $\sim$52\%, which is considerably lower than the $\sim$60\% range reported by studies dealing with a scope of a single project.
    \item Existing classification results rarely report Cohen's kappa (hence forth Kappa) metric, which accounts for cases where classification categories (a.k.a classes) are unevenly  distributed. Such cases make the accuracy metric somewhat misleading. For example, if the corrective class accounted for 98\% of the commits, and each of the remaining classes accounted for 1\% of the commits, then a simple classification model which always classified commits as corrective would have an impressive accuracy of 98\%. Its Kappa on the other hand, would be 0, making this model much less appealing.
    \item Our previous work \cite{levinIcsme2016} shows that source code change types as defined by Fluri et al. \cite{fluri2006classifying} are statistically significant in the context of maintenance activity categories defined by Mockus et al. \cite{mockus2000identifying}. We believe that boosting (i.e. increasing) the accuracy and kappa characteristics of commit classification into maintenance activities could improve the quality and accuracy of developers' maintenance profiles and the prediction models thereof \cite{levinIcsme2016} (see also section \ref{sec:discussion}).
\end{enumerate}

In this work we seek to design a commit classification model capable of providing high accuracy and Kappa across different projects. 
Our intuition is to try and capture such information, that is not unique to commits made in one project or another, but is of a rather generic nature.
Fluri's taxonomy of source code changes for object-oriented programming languages (OOPLs) \cite{fluri2006classifying} consists of 48 (47 $+$ an ''unknown type``) different change types, all of which are project agnostic and describe a meaningful action performed by a developer in a commit (e.g., \textit{statement\_delete}, \textit{statement\_insert}, \textit{removed\_class}, \textit{additional\_class} etc). Fluri's taxonomy of source code changes is therefore a nice fit for our need to capture project agnostic information pertaining to developer commits.

\begin{enumerate}[leftmargin=*,labelindent=16pt, label={\textbf{RQ. \arabic*.}}]
    \item Can source code changes be utilized to boost commit classification into maintenance activities?
    \item How does the quality of models which utilize source code changes compare to that of traditional models which use word frequency analysis?
\end{enumerate}

\section{Related Work}\label{sec:relatedWork}

Classifying commits into maintenance activities is commonly accomplished by inspecting commits' comment text and searching for indicative keywords \cite{mockus2000identifying, hindle2009automatic, levinIcsme2016, fischer2003populating, sliwerski2005changes}. Such keywords can be obtained using various techniques, such as a word-frequency analysis with normalization (e.g., stemming). Mockus et al. \cite{mockus2000identifying} was the first to employ a comment based commit classification, and reported the accuracy to be $\sim$60\% when this method was applied in the scope of a single project - a large multi-million line real-time telecommunications software system.

Recent work explored using additional information such as commits' author and module, to classify commits both within a single software project, and cross-projects \cite{hindle2009automatic}. Within a single project, the reported accuracy ranged from $\sim$35\% to 70\% (accuracy fluctuated considerably depending on the project), for cross-project classification the accuracy was $\sim$52\% \cite{hindle2009automatic}.

A slightly different technique was used by Amor et al. \cite{amor2006discriminating} who explored classifying maintenance activities in the FreeBSD project by applying a Naive Bayes classifier on commits' comments without an apparent use of keywords. The reported accuracy of classifying a random sample was $\sim$70\%. The sample's size was not specified.

A summary of the exiting results for commit classification into maintenance activities can be found in table \ref{currentResults}.
 
\begin{table}[h]    
    \center
    \renewcommand{\arraystretch}{1.2}
    \caption{Classifying commits into maintenance activities, existing results \cite{hindle2009automatic,amor2006discriminating,mockus2000identifying}}
    \label{currentResults}
    \begin{tabular}{|c|c|}        
    \hline
    \rowcolor{lightgray} \textbf{Scope} &  \textbf{Max Accuracy} \\
    \hline
    Single Project &  70\% \\
    \hline
    Cross-Project &  52\% \\
    \hline                  
    \end{tabular}    
\end{table} 
In this work we were able to achieve an accuracy of 76\% and Cohen's kappa of 63\% in the context of cross-project commit classification, an improvement of over 20 percentage points and a relative boost of $\sim$40\% in accuracy compared to previous results.

\section{Statistical Methods}\label{sec:statMethods}
\label{sec:methods}

Picking the optimal classifier for a real-world classification problem is hardly a simple task \cite{fernandez2014we}, however, Random Forest (RF) \cite{ho1998random, breiman2001random} and Gradient Boosting Machine (GBM) \cite{friedman2001greedy,rGbm,ridgeway2007generalized,caruana2006empirical,caruana2008empirical} classifiers are generally considered top performers \cite{caruana2006empirical,fernandez2014we}. 
In addition, we also use J48 \cite{frank2005weka,rWeka}, a variation of the C4.5 \cite{quinlan2014c4} algorithm. The RF and GBM models are most likely to outperform the simpler J48, but the latter, in contrast to the formers, is capable of providing a human readable representation of its decision tree. We find this ability valuable, since inspecting the decision tree can provide additional insights.
An example of a decision tree produced by the J48 classifier can be found in figure \ref{fig:j48-tree}. It is a decision tree for our keyword based commit classification model, described in section \ref{modelTypes}.

To evaluate the different commit classification models we employ common statistical measures for classification performance. For a given class ${L \in \{\operatorname{adaptive},\operatorname{corrective},\operatorname{perfective}\}}$, 
$\operatorname{TP}_L$ is the number of commits correctly classified as class $L$;
$\operatorname{FP}_L$ is the number of commits incorrectly classified as class $L$;
$\operatorname{FN}_L$ is the number of commits of class $L$ that were incorrectly classified.

\begin{itemize}[itemsep=3pt]
    \item \textit{Precision}$_L$ $ = \frac{\operatorname{TP}_L}{\operatorname{TP}_L + \operatorname{FP}_L}$, the number of commits correctly classified as class $L$, divided by the total number of commits classified as class $L$.
    \item \textit{Recall}$_L$ $ = \frac{\operatorname{TP}_L}{\operatorname{TP}_L + \operatorname{FN}_L}$, the number of commits correctly classified as class $L$, divided by the actual number of $L$ class commits in the dataset. 
    \item \textit{Accuracy} $ = \frac{\sum_{L \in \{a,c,p\}}\operatorname{TP}_L}{\sum_{L \in \{a,c,p\}} (\operatorname{TP}_L +  \operatorname{FP}_L)}$, the proportion of correctly classified commits out of all classified commits.
    \item \textit{No Information Rate (NIR)}, measures the accuracy of a trivial classifier which classifies all commits with using a single class, the one that is most frequent, in our case - corrective.
    \item \textit{Kappa} - Cohen's kappa, often considered helpful as a measure that can handle  both multi-class and imbalanced class problems (see section \ref{sec:intro}).
    \item \textit{P-Value [Accuracy $>$ NIR]}, the $p$-value for the null hypothesis that the ''Accuracy $\leq$ NIR`` (i.e., the accuracy of a given predictive model) . A low $p$-value allows one to reject the null hypothesis in favor of the alternative hypothesis that the ''Accuracy $>$ NIR``.
\end{itemize}

\section{Research Method}

Our work consists of the following main procedures:
\begin{enumerate}
    \item Select candidate software projects and harvest their commit data such as commit message and source code changes performed as part of the commit (see sections \ref{sec:selectingRepos}, \ref{sec:distillingRepos}).
    \item Create a labeled commit dataset by sampling commits and manually labeling them. Each label is a maintenance category, i.e. one of the following: corrective, perfective, or adaptive (see section \ref{sec:manualLabels}).
    \item Inspect the agreement level of the manual labeling procedure by sampling 10\% of the labeled dataset created by the first author, and have the second author independently label it (see section \ref{sec:manualAgreement}).
    \item Devise predictive models that utilize source code changes for the task of commit classification into maintenance activities (see section \ref{modelTypes}).
    \item Evaluate the devised models using two mutually exclusive datasets obtained by splitting the labeled dataset into
    \begin{enumerate*}[label={(\arabic*)}]
        \item a \textit{training} dataset, consisting of 85\% of the labeled dataset, and
        \item a \textit{test} dataset, consisting of the remaining 15\% of the labeled dataset which was never employed as part of the training process
    \end{enumerate*}
    (see section \ref{sec:evaluaton}).
\end{enumerate}

\section{Data Collection}\label{sec:dataCollection}

\subsection{Selecting candidate software projects}\label{sec:selectingRepos}

We use GitHub \cite{gitHub} as the data source for this work due to its popularity and rich query options.
Candidate repositories were selected according to the following criteria, which we designed to target data-rich repositories:
\begin{itemize}
    \item Used the Java programming language
    \item Had more than 100 stars (i.e. more than 100 users have ''liked`` these repositories)
    \item Had more than 60 forks (i.e., more than 60 users have ''copied`` these repositories for their own use)
    \item Had their code updated since 2016-01-01 (i.e., these repositories are active)
    \item Were created before 2015-01-01  (i.e., these repositories have been around)
    \item Had size over 2MB (i.e. these repositories are of considerable size)
\end{itemize}

Out of all candidates we selected 11 projects which are well known in the open source community, and cover a wide range of software domains such as IDEs, programming languages (that were implemented in Java), distributed database and storage platforms, and integration frameworks. 
\begin{enumerate}
    \item \textbf{RxJava} - a library for composing asynchronous and event-based programs for the Java VM.
    \item \textbf{Intellij Community Edition} - A popular IDE for the Java programming language.
    \item \textbf{HBase} -  A distributed, scalable, big data store.
    \item \textbf{Drools} -  A Business Rules Management System solution.
    \item \textbf{Kotlin} - A Statically typed programming language for the JVM, Android and the browser by JetBrains.
    \item \textbf{Hadoop} - A framework that allows for the distributed processing of large data sets across clusters of computers.
    \item \textbf{Elasticsearch} - A distributed search and analytics engine.
    \item \textbf{Restlet} - A RESTful web API framework for Java.
    \item \textbf{OrientDB} - A Distributed Graph Database with the flexibility of Documents in one product.
    \item \textbf{Camel} - An open source integration framework based on known Enterprise Integration Patterns.
    \item \textbf{Spring Framework} - An application framework and inversion of control container for the Java platform.
\end{enumerate}

\subsection{Distilling source code changes}\label{sec:distillingRepos}

The source code changes distilling was carried out using a designated VCS mining platform we have built on top of Spark \cite{zaharia2010spark, sparkSite}, a state of the art framework for large data processing.

After downloading (cloning) the repositories from GitHub, for each repository $r$ where $1\leq r \leq 11$ we created a series of patch files $\{p_{i}^{r}\}_{i=1}^{N_r}$, where $N_r$ is the latest revision number for repository $r$. Each patch file $p_{i}^{r}$ was responsible for transforming repository $r$ from revision $r_{i-1}$ to revision $r_{i}$, where $r_0$ is the empty repository. By initially setting repository $r$ to revision $1$ (i.e. the initial revision) and then applying all patches ${\{p_{i}^{r}\}}_{i=2}^{N_r}$ in a sequential manner, the revision history for that repository was essentially replayed. Conceptually, this was equivalent to the case of all developers performing their commits sequentially one by one according to their chronological order.

To distill source code changes as per the taxonomy defined by Fluri et al., we repeatedly applied ChangeDistiller (\cite{gall2009change, fluri2008discovering, martinez2013automatically,fluri2007change}) on every two consecutive revisions of every Java file in every repository we had selected to be part of the dataset, essentially extracting source code changes from the entire project's history.

\subsection{Creating a labeled commit dataset}\label{sec:manualLabels}

The first author \textit{manually} classified a randomly sampled set of $\sim$100 commits from each of the studied 11 repositories.
To improve classification quality the projects' issue tracking systems (e.g. JIRA by Atlassian \cite{jira}) was often used. The JIRA contained the tickets occasionally referenced in developers' commits. Such tickets (a.k.a. issues) typically contain additional information about the feature or bug the referencing commit was trying to address. Moreover, tickets sometimes had their own classification categories such as ''feature request``, ''bug``, ''improvement`` etc., but unfortunately they were not very reliable as developers were not always consistent with their categories. For instance, in some cases bug fixes were labeled as ''improvement``, and while fixing a bug is indeed an improvement, according to the classification categories we use (Mocuks et. al. \cite{mockus2000identifying}), bug fixes should be classified corrective while improvements should be classified perfective. Some developers used the term ''fix`` even when they referenced feature requests, e.g. ''fixed issue \#N``, where ''issue \#N`` spoke of a new feature or an improvement that did not necessarily report a bug. These observations are consistent with Herzig et al \cite{herzig2013s} who reported that 33.8\% of the bug reports they studied were misclassified.

In cases where the lack of supporting information (e.g., in-descriptive ticket and / or commit message) prevented us from classifying a certain commit with satisfactory confidence, that commit was dropped from the dataset and replaced by a new one, selected randomly from the same project repository. If we were unable to classify the replacement commit as well, we would repeat this routine until we found a commit that we were able to confidently classify.
Further rules of thumb we used for classifying were as follows:
\begin{itemize}
    \item Javadoc and comment updates were considered perfective
    \item Fixing a broken unit test or build was considered corrective
    \item Adding new unit test(s) was considered perfective
    \item Performance improvements that resulted from an open ticket in the issue tracking system were considered corrective
    \item Performance improvements that did NOT result from an open ticket in the issue tracking system were considered perfective
\end{itemize}

We made efforts to prevent class starvation (i.e., not having enough instances of a certain class) which could in turn substantially degrade models' performance, and in case we detected a considerable imbalance in some project's classification categories (a.k.a classes), we added more commits of the starved class from the same project. This balancing was done by repeatedly sampling and manually classifying commits until a commit of the starved class was found.
The final dataset consisted of 1151 manually classified commits and was made open access \cite{stanislav_levin_2017_835534}. This dataset contained 100-115 from each project. 
43.4\% (500 instances) were corrective, 35\% (404 instances) were perfective, and 21.4\% (247 instances) were adaptive. These commits yielded 33,149 source code changes.

\subsection{Inspecting manual labeling agreement}
\label{sec:manualAgreement}

In order to inspect manual classification agreement, we randomly selected 110 commits out of the 1151 commits that had been labeled by the first author, 10 random commits from each of the 11 projects, and had these commits independently labeled again by the second author.

At first the agreement stood at 79\%. After discussing the conflicts and sharing the guidelines used by the first author in more detail, the agreement level rose to 94.5\%. According to the one sample proportion test \cite{altman1990practical}, the error margin for our observed agreement level was 4.2\%, and the estimated asymptotic 95\% confidence interval was [90.3\%, 98.7\%]. 
This indicates that both authors were in agreement about the labels for the vast majority of cases once they employed the same guidelines (see section \ref{sec:manualLabels}).

\section{Commit Classification Models}
\label{modelTypes}
\raggedbottom

We split the labeled dataset into a training dataset and a test dataset, 85\% and 15\% respectively.
The model training phase consists of using 5 time repeated 10-fold validation for each compound model (which boils down to performing a 10-fold cross validation process 5 different times and averaging the results). Then, the trained models are evaluated using the test dataset - the 15\% split that did not take part in the model training process.
Our statistical computations are carried out in the R statistical environment \cite{R}, where we extensively use the R caret package \cite{rCaret} for the purpose of model training and evaluation.

\subsection{Utilizing word frequency analysis}

First we classified the test dataset (the 15\% of the entire labeled dataset) using a naive method to set an initial baseline.
The naive method is based solely on searching for pre-defined words gathered from previous work \cite{levinIcsme2016}, and returning the most frequent class (i.e., corrective) in case none of the keywords were present in a commit's message, see table \ref{tab:veryNaiveClassification} for more details. The results showed that 34.8\% of the commits in the test dataset (60 commits) did not have any of the keywords present in their commit message, and were therefore automatically classified corrective. In addition, we can note the low recall of the perfective class, as opposed to the high recall of the corrective class (which accounts for most of the commits in the classified dataset).

\begin{table}
    \center
    \renewcommand{\arraystretch}{1.2}
        \caption{Naive model's confusion matrix}        
        \small
        \resizebox{.47\textwidth}{!}{
        \begin{tabular}{|c|c|c|c|}        
        \hline
        \rowcolor{lightgray} \backslashbox{classified as}{true class} & \textbf{adaptive} &  \textbf{corrective} &  \textbf{perfective} \\
        \hline
        \textbf{adaptive} &  \textbf{18}  & 2 & 16 \\
        \hline
        \textbf{corrective}  &   18 & \textbf{72} & 37 \\
        \hline              
        \textbf{perfective} &     1 &  1 & \textbf{7} \\
        \hline        
        \Xhline{1pt}                    
        \multicolumn{1}{r|}{\textbf{Recall:}} & 48\% & 96\% & 11\% \\ \cline{2-4}
        \multicolumn{1}{r|}{\textbf{Precision:}} & 50\% & 56\% & 77\% \\ \cline{2-4}
        \multicolumn{1}{r|}{\textbf{Accuracy:}} & \multicolumn{3}{c|}{56\%} \\ \cline{2-4}
        \multicolumn{1}{r|}{\textbf{Kappa:}} & \multicolumn{3}{c|}{29\%} \\ \cline{2-4}
        \multicolumn{1}{r|}{\textbf{No Information Rate (NIR):}} & \multicolumn{3}{c|}{ 43\% } \\ \cline{2-4}
        \multicolumn{1}{r|}{\textbf{P-Value [Accuracy $>$ NIR]:}} & \multicolumn{3}{c|}{ 0.0005 }  \\ \cline{2-4}
        \end{tabular}}
        \label{tab:veryNaiveClassification}
\end{table}    

Due to the high percentage of commits without any of the keywords we had defined, we then tried to fine-tune the keywords we search for.
We performed an additional experiment using the same classification method, only this time the keywords were obtained by employing a word frequency analysis with normalization of the commit messages. This time 28\% of the commits did not have any of the keywords present in their commit message. These findings led us to believe that the number of commits having none of the keywords used by keyword based classification models is quite considerable.

\subsection{Utilizing source code changes (RQ. 1)}\label{sec:models}

The subject of dealing with missing values in a classification problem is broadly covered by Saar-Tsechansky et al. \cite{saar2007handling}, who describe two common methods employed to overcome this issue: 
\begin{enumerate*}
    \item imputation, where the missing values are estimated from the data that are present, and
    \item reduced-feature models, which employ only those features that will be known for a particular test case (i.e., only a subset of the features that are available for the entire training dataset), so that imputation is not necessary.
\end{enumerate*}
Since our dataset consists of two different data types, keywords and source code changes, we use reduce-feature models, which are reported to outperform imputation \cite{saar2007handling} and represent our use-case more naturally.
In addition, since the missing feature patterns in our dataset are known in advance, i.e., given a commit only the keywords can be missing, its source code changes are always present, we can pre-compute and store two models; one to be used when all features are present (keywords $+$ source code changes), and the other when only a subset is available (source code changes only).
We define the notion of a \textbf{compound} model (similarly to the ''classifier lattice`` \cite{saar2007handling}) which uses two separate models for classifying commits with, and without (pre-defined) keywords in their commit message.
The classify routine of the compound model is pseudo-coded in listing \ref{compound-classify-cod}.

\begin{minipage}{0.95\linewidth}
    \begin{lstlisting}[label=compound-classify-cod,breaklines,mathescape=true,caption=Compound model's classify routine,frame=tb]
classify(commit) {
    if(hasKeywords(commit.comment)) { 
        classifyWith($\operatorname{model}_{KW}$,commit);
    } else {
        classifyWith($\operatorname{model}_{\overline{KW}}$,commit);
    }
}
    \end{lstlisting}
    \vspace{1em}
\end{minipage}

\noindent Given a commit $C$, the compound model first checks if $C$'s commit message has any keywords, if so, the model defined as $\operatorname{model}_{KW}$ is used to classify $C$, otherwise (i.e., no keywords found in $C$'s commit message), the model defined as $\operatorname{model}_{\overline{KW}}$ is used to classify $C$.
Each of the models $\operatorname{model}_{KW}$ and $\operatorname{model}_{\overline{KW}}$ may or may not be a reduced-feature model, depending on whether it employs the full set of features (both keywords and source code changes), or only a subset of it (either keywords or source code changes).\\
We define $\operatorname{model}_{\overline{KW}}$ and $\operatorname{model}_{KW}$ to be one of the following model types:

\begin{itemize}
    \item \underline{Keywords} model, which relies solely on keywords to classify commits. 
        \label{keywordsExtraction}
        The features used by this model are keywords obtained by performing the following transformations on the commit comment fields:
        
        \begin{enumerate}
            \item Removed special characters 
            \item Made lower case (case-folding)
            \item Removed English stopwords
            \item Removed punctuation
            \item Striped white-spaces
            \item Performed stemming
            \item Adjusted frequencies so that each comment can contribute a given word only once
            \item Removed custom words such as developer names, projects names, VCSs lingo (e.g., head, patch, svn, trunk, commit), domain specific terms (e.g., http, node, client):
            {\small ''patch``, ''hbase``, ''checksum``, ''code``, ''version``, ''byte``, ''data``, ''hfile``, ''region``, ''schedul``, ''singl``, ''can``, ''yarn``, ''contribut``, ''commit``, ''merg``, ''make``, ''trunk``, ''hadoop``, ''svn``, ''ignoreancestri``, ''node``, ''also``, ''client``, ''hdfs``, ''mapreduc``, ''lipcon``, ''idea``, ''common``, ''file``, ''ideadev``, ''plugin``, ''project``, ''modul``, ''find``, ''border``, ''addit``, ''changeutilencod``, ''clickabl``, ''color``, ''column``, ''cach``, ''jbrule``, ''drool``, ''coprocessor``, ''regionserv``, ''scan``, ''resourcemanag``, ''cherri``, ''gong``, ''ryza``, ''sandi``, ''xuan``, ''token``, ''contain``, ''shen``, ''todd``, ''zhiji``, ''tan``, ''wangda``, ''timelin``, ''app``, ''kasha``, ''kashacherri``, ''messag``, ''spr``, ''camel``, ''http``, ''now``, ''class``, ''default``, ''pick``, ''via``.}

            \item We then selected the 10 most frequent words from each of the three maintenance activities in the test dataset:                             
                \begin{itemize}
                    \item Corrective:
                        \begin{enumerate*}[label={(\arabic*)},before=\itshape,font=\normalfont]
                            \item fix
                            \item test
                            \item issu
                            \item use
                            \item fail
                            \item bug
                            \item report
                            \item set 
                            \item error 
                            \item npe
                        \end{enumerate*}
                    \item Perfective:
                        \begin{enumerate*}[label={(\arabic*)},before=\itshape,font=\normalfont]
                            \item test
                            \item remov
                            \item use
                            \item fix
                            \item refactor
                            \item method
                            \item chang 
                            \item add
                            \item improv
                            \item new                           
                        \end{enumerate*}                    
                    \item Adaptive:
                        \begin{enumerate*}[label={(\arabic*)},before=\itshape,font=\normalfont]
                            \item support
                            \item add
                            \item implement
                            \item new
                            \item allow
                            \item use
                            \item method
                            \item test
                            \item set
                            \item chang
                        \end{enumerate*}
                \end{itemize}                        
                        
            It can be seen that some of the words (as obtained by our commit message word frequency analysis) overlap between categories. The words \textit{''test``} and \textit{''use``} appear in all three categories; the word \textit{''fix``} appears in both the corrective and perfective categories; the words \textit{''method``}, \textit{''chang``}, \textit{''add``} and \textit{''new``} appear both in the perfective and adaptive categories; and the word \textit{''set``} appears both in the corrective and adaptive categories.
            These word overlaps may indicate that keywords alone are insufficient to accurately classify commits into maintenance activities, and need to be augmented with additional information in order to improve classification accuracy.
            
            For the purpose of building the Keywords model type, we remove multiple occurrences of the same word (so that each word appears only once per maintenance category) and remain with the following set of words:
            \begin{enumerate*}[label={(\arabic*)},font=\normalfont]
                \item add
                \item allow
                \item bug
                \item chang
                \item error
                \item fail
                \item fix
                \item implement
                \item improv
                \item issu
                \item method
                \item new
                \item npe
                \item refactor
                \item remov
                \item report
                \item set
                \item support
                \item test
                \item use.
            \end{enumerate*}        
        \end{enumerate}
    \item \underline{(Source Code) Changes} based model, which relies solely on source code changes to classify commits.
        The features used by this model are source code change types \cite{fluri2006classifying} obtained by distilling commits, as described earlier in this section.
    \item \underline{Combined} (Keyword + Source Code Change Types) model, which uses both keywords and source code change types to classify commits.
        The features used by this type of models consist of both keywords and source code change types. 
\end{itemize}

\noindent A summary of the model components can be found in table \ref{internalComponentTypes}.

\begin{table}[h]    
    \center
    \renewcommand{\arraystretch}{1.2}
    \caption{Reduced-feature model components}
    \label{internalComponentTypes}
    \begin{tabular}{|c|c|}        
    \hline
    \rowcolor{lightgray} \textbf{Model Type} &  \textbf{Model Features} \\
    \hline
    Keywords &  Words \\
    \hline
    Changes  & Source Code Change Types  \\
    \hline              
    Combined & Words + Source Code Change Types  \\
    \hline
    \end{tabular}    
\end{table}

For example, a commit where two methods were added (source code change type ''additional\_functionality``), and one statement was updated (source code change type ''statement\_updated``) and has a commit message that says ''Refactored blob logic into separate methods`` will be treated differently by each of the model types indicated in table \ref{internalComponentTypes}.\\
The Keywords model extracts features represented by tuples of size 20, and given the commit above would extract the following features:
$\overbrace{(0\dots1\dots1\dots0)}^\text{20}$ with ``1'' in the coordinates that represent the words \textit{''refactor``} and \textit{''method``}. The count of each keyword is at most one, i.e., duplicate keywords are counted only once. Source code changes are ignored, since the Keywords model type does not consider source code changes.\\
The Changes model extracts features represented by tuples of size 48 (since there are 48 different source code change types), and given the commit above would extract the following features:
$\overbrace{(0\dots2\dots1\dots0)}^\text{48}$ with ''2`` in the coordinate that represents the source code change type \textit{''additional\_functionality``} and ``1'' in the coordinate that represents \textit{''statement\_updated``}. In contrast to the case of the Keywords model, all occurrences of every source code change type are counted in. Keywords in the commit message are ignored, since the Changes model type does not consider keywords.\\
The Combined model extracts features represented by tuples of size 68 ($=$ 48 source code change types + 20 keywords), and given the commit above would extract the following features:
$\overbrace{(\underbrace{0\dots1\dots1\dots0}_\text{20}\underbrace{\dots0\dots2\dots1\dots0}_\text{48})}^\text{68}$, with ''2`` in the coordinate that represents the source code change type \textit{''additional\_functionality``}, and ''1`` in the coordinates that represent the source code change type \textit{''statement\_updated``}, the keyword \textit{''refactor``}, and the keyword \textit{''method``}. The Combined model type captures both keywords and source code change types - hence its name.

In the next sections we evaluate and compare different compound models by considering the different combinations of their $\operatorname{model}_{KW}$ and $\operatorname{model}_{\overline{KW}}$ model components.
The evaluation process consists of the following steps:
\begin{enumerate}
    \item Select the model component $\operatorname{model}_{KW}$
    \item Select the model component $\operatorname{model}_{\overline{KW}}$    
    \item Select an underlying classification algorithm for the compound model, which determines the algorithm to be used by each of the model components $\operatorname{Model}_{KW}$ and $\operatorname{Model}_{\overline{KW}}$ (J48, GBM, or RF see section \ref{sec:methods}).        
\end{enumerate}

\section{Results (RQ. 2)}\label{sec:evaluaton}

\begin{table}
    \renewcommand{\arraystretch}{1.2}
    \centering
    \caption{Training dataset compound models performance}
    \label{tab:model_combos}
    \begin{tabular}{|p{0.13\linewidth}|c|c|c|c|}
        \hline
        \rowcolor{lightgray} \centering \textbf{Alg.}  & $\operatorname{Model}_{KW}$ & $\operatorname{Model}_{\overline{KW}}$ & \textbf{Accuracy} & \textbf{Kappa} \\
        \hline
         \centering  \multirow{9}{*}{J48}  & \multicolumn{2}{c|}{Combined}                                                      &      69.0\% & 51.7\% \\
         \cline{2-5}
         & Combined     &       Keywords                                                                            &        67.7\% & 50.2\%  \\
         \cline{2-5}
         & Combined     &       Changes                                                                             &         69.2\% & 51.9\% \\
         \cline{2-5}
         \Xcline{2-5}{1.5pt}
         &  Keywords    &     Combined                                                        &    69.8\% &  53\% \\
         \cline{2-5}
         & \multicolumn{2}{c|}{\cellcolor{orange} Keywords}                                                                        &       \cellcolor{orange} 68.5\% & \cellcolor{orange} 51.5\% \\
         \cline{2-5}
         & \cellcolor{lime} Keywords    &  \cellcolor{lime} Changes  &    \cellcolor{lime}  69.9\% & \cellcolor{lime} 53.2\% \\
         \cline{2-5}
         \Xcline{2-5}{1.5pt}
         & Changes  &       Combined                                                                                &         48.7\% & 20.1\% \\
         \cline{2-5}         
         & Changes  &       Keywords                                                                            &       47.4\% & 17.2\%  \\
         \cline{2-5}
         & \multicolumn{2}{c|}{Changes}                                                                         &       48.8\% & 18.6\% \\
        \Xhline{3pt}

         \multirow{9}{*}{\parbox{\linewidth}{\centering GBM}}  & \multicolumn{2}{c|}{\cellcolor{lime}Combined}  & \cellcolor{lime} 72.0\% &  \cellcolor{lime} 56.2\% \\
         \cline{2-5}
         & Combined   &         Keywords                                                                            &      69.0\% & 51.8\% \\
          \cline{2-5}
         &   Combined   &     Changes                                                                               &   72.0\% & 55.9\% \\
          \cline{2-5}
          \Xcline{2-5}{1.5pt}
         & Keywords   &        Combined                                                                             &       71.6\% & 56.0\% \\
          \cline{2-5} 
          
         & \multicolumn{2}{c|}{\cellcolor{orange} Keywords}                                                                        & \cellcolor{orange} 68.5\% & \cellcolor{orange} 51.4\% \\         
          \cline{2-5}
         & Keywords    &       Changes                                                                          &       71.5\% & 55.6 \\
          \cline{2-5}
          \Xcline{2-5}{1.5pt}
         & Changes  &       Combined                                                                                &       54.1\% & 26.9\% \\
          \cline{2-5}
         & Changes   &      Keywords                                                                            &      51.0\% & 22.4\% \\
          \cline{2-5}
         & \multicolumn{2}{c|}{Changes}                                                                         &       54.3\% & 26.9\% \\
         \Xhline{3pt}

        \multirow{9}{*}{\parbox{\linewidth}{\centering RF}}                     &  \multicolumn{2}{c|}{Combined}    &       73.1\% & 57.8\% \\
          \cline{2-5}
          & Combined     &       Keywords                                                                           &      69.5\% & 52.6\% \\
          \cline{2-5}
          & Combined     &       Changes                                                                            &      71.9\% & 55.7\% \\
          \cline{2-5}
          \Xcline{2-5}{1.5pt}
          & Keywords    &       Changes                                                                         &       72.2\% & 56.4\% \\
          \cline{2-5}
          & \cellcolor{lime} Keywords    &   \cellcolor{lime}    Combined       &    \cellcolor{lime}  73.6\% & \cellcolor{lime} 58.9\% \\
          \cline{2-5}
          & \multicolumn{2}{c|}{\cellcolor{orange} Keywords}                                                                       & \cellcolor{orange} 69.8\% & \cellcolor{orange} 53.4\% \\
          \cline{2-5}        
          \Xcline{2-5}{1.5pt}
          & Changes  &       Combined                                                                               &       54.5\% & 26.6\% \\
          \cline{2-5}
          & Changes  &       Keywords                                                                           &       50.6\% & 21.1\% \\
          \cline{2-5}
          & \multicolumn{2}{c|}{Changes}                                                                        &       52.9\% & 23.4\% \\
    \hline         
    \end{tabular}
\end{table}

Table \ref{tab:model_combos} describes an exhaustive set of combinations for selecting the pair of $(\operatorname{Model}_{KW}, \operatorname{Model}_{\overline{KW}})$ models, given that each can be one of the three model types defined in table \ref{internalComponentTypes}.
Each row in table \ref{tab:model_combos} represents a compound model, defined by the selection of $(\operatorname{Model}_{KW}, \operatorname{Model}_{\overline{KW}})$. The classification accuracy and Kappa achieved by a given compound model are reported in the corresponding Accuracy and Kappa columns.
The best performing compound model for each classification algorithm is highlighted in lime-green, and the keywords based model (where both $\operatorname{Model}_{KW}$ and , $\operatorname{Model}_{\overline{KW}}$ are of the Keywords model type) is highlighted in orange so that it can be easily compared to compound models that utilize source code changes.\\

Following our main research questions (see section \ref{sec:intro}), the accuracy and Kappa results for each compound model during the training (see table \ref{tab:model_combos}) reveal that the compound models that use either \fbox{$ \displaystyle \operatorname{Model}_{\overline{KW}} = \operatorname{Combined}$} or \fbox{$ \displaystyle \operatorname{Model}_{\overline{KW}} = \operatorname{Changes}$} achieve higher accuracy and Kappa when compared to models with the same $\displaystyle \operatorname{Model}_{KW}$ component but  with ${\operatorname{Model}_{\overline{KW}} = \operatorname{Keywords}}$, regardless of the underlying classification algorithm (J48, GBM or RF). This comes as no surprise, as one could expect keyword based models would have trouble accurately classifying commits that do not have any keywords in their commit message. Table \ref{tab:model_combos} also reveals that models that rely solely on commit messages have higher accuracy and kappa than models that rely solely on source code changes (under all three algorithms).

Further accuracy and Kappa statistics pertaining to the training stage of the best performing model for each algorithm can be found in table \ref{resamplesStatsAccuracy} and table \ref{resamplesStatsKappa} respectively. From table \ref{resamplesStatsAccuracy} and table \ref{resamplesStatsKappa} we can learn that during the training stage, the RF model consistently outperforms the J48 and even the GBM model, in both accuracy and Kappa, across all of the cuts: minimum, 1-st quartile (25-th percentile), median, mean, 3-rd quartile (75-th percentile) and maximum. In particular, the minimum accuracy and Kappa of the RF are notably higher than its competitors.

\begin{table}
    \center
    \newcommand\Tstrut{\rule{0pt}{2.6ex}}
    \renewcommand{\arraystretch}{1.2}
    \setlength\extrarowheight{2pt}
    \caption{Training dataset accuracy, best model per algorithm}
    \label{resamplesStatsAccuracy}
    \small
    \begin{tabular}{|c|c|c|c|c|c|c|}
    \hline
    \rowcolor{lightgray} Alg. & Min. & 1-st Q. & Median & Mean & 3-rd Q. &  Max. \\
    \hline
    J48 & 60.8\% & 66.4\% & 70.1\% & 69.9\% & 73.4\% & 80.6\% \\
    \hline
    GBM & 60.8\% & 69.2\% & 72.1\% & 72.0\% & 75.2\% & 80.8\% \\
    \hline              
    \rowcolor{lime} RF &  65.6\% &  70.4\%  & 73.4\% & 73.6\%  & 76.6\% & 82.8\% \\
    \hline
    \end{tabular}    
\end{table}

\begin{table}
    \center
    \renewcommand{\arraystretch}{1.2}
    \setlength\extrarowheight{2pt}
    \caption{Training dataset Kappa, best model per algorithm}
    \label{resamplesStatsKappa}
    \small
    \begin{tabular}{|c|c|c|c|c|c|c|}
    \hline
    \rowcolor{lightgray} Alg. & Min. & 1-st Q. & Median & Mean & 3-rd Q. &  Max. \\
    \hline
    J48 & 38.4\% & 47.9\% & 53.4\% & 53.2\% & 58.8\% & 69.7\% \\
    \hline
    GBM & 38.3\% & 51.8\% & 56.9\% & 56.2\% & 60.6\% & 70.0\% \\
    \hline              
    \rowcolor{lime} RF &  45.5\% & 54.1\% & 58.6\% & 58.9\% & 63.3\% & 73.5\% \\
    \hline
    \end{tabular}    
\end{table}

The top performing models were then used to classify the test dataset, consisting of 15\% of the entire labeled dataset, see table \ref{tab:testsetResults}.
The ultimate winner was the RandomForest compound model with $\operatorname{Model}_{KW}=\operatorname{Keywords}$ and $\operatorname{Model}_{\overline{KW}}=\operatorname{Combined}$.
A detailed confusion matrix for this champion  model can be found in table \ref{tab:bestClassification}.


\begin{table}[h]
    \renewcommand{\arraystretch}{1.2}
    \centering
    \caption{Test dataset classification performance}
    \label{tab:testsetResults}
    \small
    \begin{tabular}{|c|c|c|c|c|}
        \hline
        \rowcolor{lightgray} \textbf{Algorithm}  & $\operatorname{Model}_{KW}$ & $\operatorname{Model}_{\overline{KW}}$ & \textbf{Accuracy} & \textbf{Kappa}  \\
        \hline
         J48 & Keywords & Changes & 70.3\% & 53.9\% \\
         \hline
         GBM & \multicolumn{2}{c|}{Combined}  & 72.6\% & 57.2\% \\
         \hline
         \rowcolor{lime} RF & Keywords & Combined & 76.7\% & 63.5\% \\
         \hline
     \end{tabular}
 \end{table}

\begin{table}
    \center
    \renewcommand{\arraystretch}{1.2}
        \caption{Keywords-Combined RF compound model's confusion matrix for the test dataset}
        \small
        \begin{tabular}{|c|c|c|c|}        
        \hline
        \rowcolor{lightgray} \backslashbox{classified as}{true class} & \textbf{adaptive} &  \textbf{corrective} &  \textbf{perfective} \\
        \hline
        \textbf{adaptive} &  \textbf{28}  & 5 & 5 \\
        \hline
        \textbf{corrective}  &   6 & \textbf{63} &  14 \\
        \hline              
        \textbf{perfective} &     3 &  7  & \textbf{41} \\
        \hline
        \Xhline{1pt}
        \multicolumn{1}{r|}{\textbf{Recall:}} & 75\% & 84\% & 68\% \\ \cline{2-4}        
        \multicolumn{1}{r|}{\textbf{Precision:}} & 73\% & 75\% & 80\% \\ \cline{2-4}        
        \multicolumn{1}{r|}{\textbf{Accuracy:}} & \multicolumn{3}{c|}{76\%} \\\cline{2-4}
        \multicolumn{1}{r|}{\textbf{Kappa:}} & \multicolumn{3}{c|}{63\%} \\\cline{2-4}
        \multicolumn{1}{r|}{\textbf{No Information Rate (NIR):}} & \multicolumn{3}{c|}{43\%} \\\cline{2-4}
        \multicolumn{1}{r|}{\textbf{P-Value [Accuracy $>$ NIR]:}} & \multicolumn{3}{c|}{$<2e^{-16}$ } \\ \cline{2-4}         
        \end{tabular}    
        \label{tab:bestClassification}
\end{table}    

\begin{figure*}
    \caption{A J48 Keywords model type (''a`` stands for adaptive, ''c`` for corrective, and ''p`` for perfective)}
    \label{fig:j48-tree}
    \includegraphics[width=1.189\textwidth,height=\textwidth,angle=90]{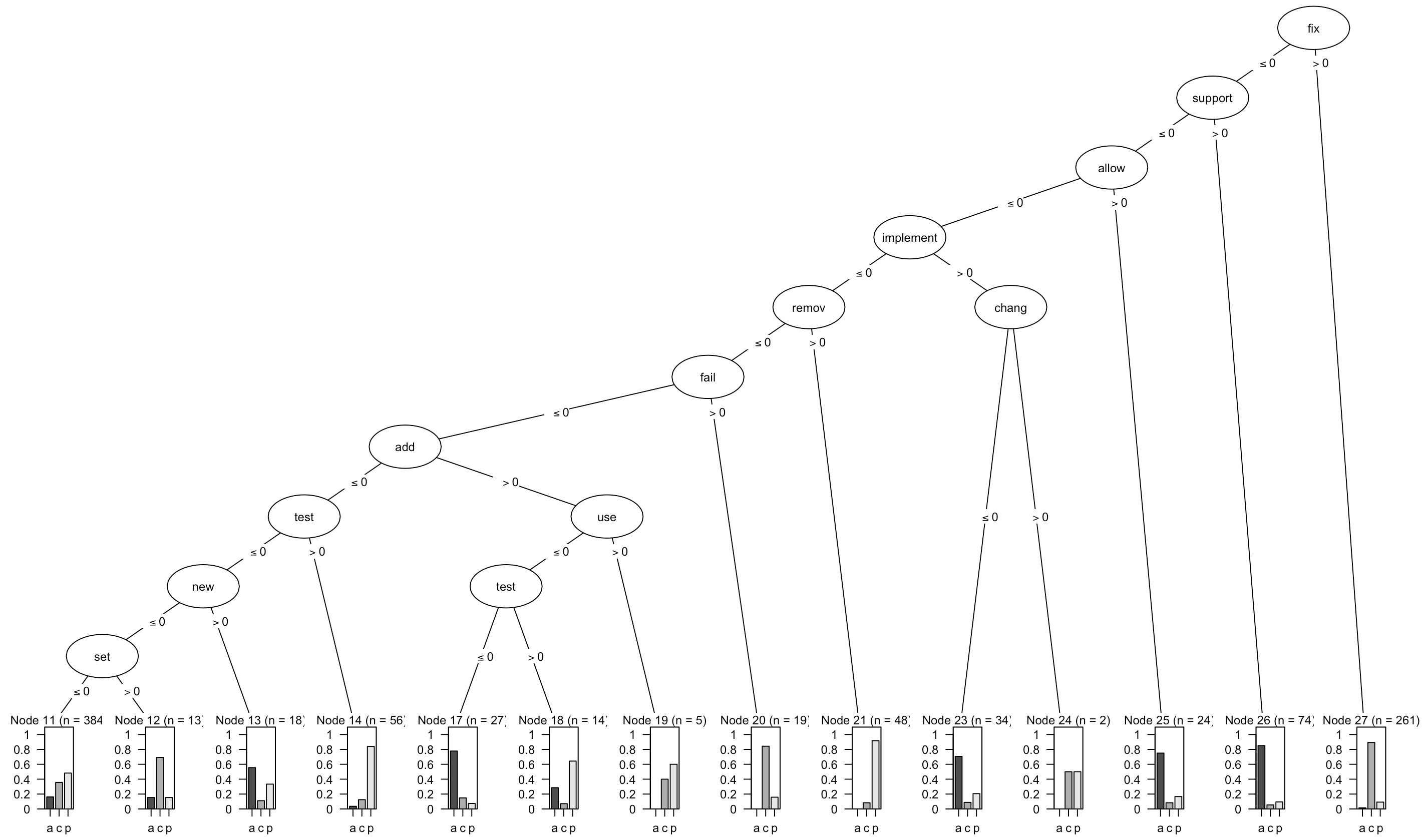}
\end{figure*}

The decision tree built by the J48 algorithm for our keyword based model (see figure \ref{fig:j48-tree}) provides some interesting insights regarding its classification process.
The word ''fix`` is the single most indicative word of corrective commits, which aligns well with our intuition, according to which commits that fix faults are likely to include the ''fix`` noun or verb in the commit message. Given that ''fix`` did not appear, the words ''support`` and ''allow`` are most indicative of adaptive commits, presumably these words are used by developers to indicate the support of a new feature, or the fact that something new is now ''allowed`` in the system. 
The combination ''implement chang`` (stemmed), given that ''fix``, ''support`` and ''allow`` did not appear, is very indicative of either perfective or corrective commits, if however, ''implement`` is not accompanied by the word ''chang`` (stemmed), the commit is likely to be adaptive.
The (stemmed) word ''remov``, given that the words ''fix``, ''support``, ''allow`` and ''implement`` did not appear, is very indicative of perfective commits, perhaps because developers often use it to describe a modification where they remove an obsolete mechanism in favor of a new one.

We also visualized the maintenance categories keyword (see section \ref{sec:models}) frequency using a word-cloud{\renewcommand{\footnotesize}{\small} \footnote{https://github.com/staslev/paper-resources/raw/promisedata-2017/Boosting-Automatic-Commit-Classification-Into-Maintenance-Activities-By-Utilizing-Source-Code-Changes/word-cloud.png}}, which revealed that the word ''test`` is particularly common in perfective commits, but is generally common in all three maintenance activity types. The word ''use`` is also common in all three maintenance activity types, but is particularly frequent in the perfective category.
The words ''fix``, ''remov`` and ''support`` are quite distinctive of their corresponding maintenance activity types: corrective, perfective and adaptive categories (respectively). 
The word ''add`` is common in adaptive commits, as well as ''allow``.

Similarly, the source code changes frequencies can also be visualized using a source-code-change-type-cloud{\renewcommand{\footnotesize}{\small} \footnote{https://raw.githubusercontent.com/staslev/paper-resources/promisedata-2017/Boosting-Automatic-Commit-Classification-Into-Maintenance-Activities-By-Utilizing-Source-Code-Changes/change-cloud.png}} which reveals that statement related changes, e.g., ''statement\_insert``, ''statement\_update`` and ''statement\_delete`` are the most common change types in all three maintenance activities (corrective, perfective, adaptive). The semantic change type ''additional\_functionality`` is common in both perfective and adaptive commits, but less so in corrective commits.

The term-cloud and J48 keyword based decision tree visualizations provide an intuition for why J48 is likely to outperform a simple word-frequency based classification. In contrast to the word-cloud, which provides ''flat`` frequencies, the J48 is capable of capturing information pertaining to the presence of multiple keywords in the same commit message, as indicated by the decision tree.
In addition, the predictor importance analysis for the champion RF model (omitted for brevity), shows numerous change types rank high, confirming their viability for the classification process.

\section{Discussion \& Applications}\label{sec:discussion}

\noindent\textbf{Improving developer's maintenance profile accuracy.} Our previous work \cite{levinIcsme2016} suggested the notion of a developer's maintenance profile, which describes the amount of commits a given developer made in each of the maintenance categories (corrective, perfective, adaptive). The models we devised in order to predict developers maintenance profiles could benefit from a more accurate classification of commits into maintenance activities as part of their training stage, possibly yielding higher prediction quality.

\noindent\textbf{Identifying anomalies in development process.} 
    The manager of a large software project should aim to control and manage its maintenance activity profiles, i.e., the volume of commits made in each maintenance activity. Monitoring for unexpected spikes in maintenance activity profiles and investigating the reasons (root cause) behind them would assist managers and other stakeholders to plan ahead and identify areas that require additional resource allocation. 
    For example, lower corrective profiles could imply that developers are neglecting bug fixing. Higher corrective profiles could imply an excessive bug count. Finding the root cause in cases of significant deviations from predicted values may reveal essential issues whose removal can improve projects' health. Similarly, exceptionally well performing projects can also be a good subject for investigation in order to identify positive patterns.
    
\noindent\textbf{Improving development team's composition.}
    Building a successful software team is hardly a trivial task as it involves a delicate balance between technological and human aspects \cite{gorla2004should, guinan1998enabling}. We believe that by using commit classification it would be possible to build reliable developer maintenance activity profiles \cite{levinIcsme2016} which could assist in composing balanced teams. We conjecture that composing a team that heavily favors a particular maintenance activity (e.g. adaptive) over the others could lead to an unbalanced development process and adversely affect the team's ability to meet typical requirements such as developing a sustainable number of product features, adhering to quality standards, and minimizing technical debt so as to facilitate future changes.

\section{Threats to validity}
\label{sec:threatsToValidity}

\noindent\textbf{\textit{Threats to Statistical Conclusion Validity}} are the degree to which conclusions about the relationship among variables based on the data are reasonable. \\
Our results are based on manually classifying 1151 commits, over 100 commits from each of the studied 11 projects.
The projects originated from various professional domains such as IDEs, programming languages, distributed database and storage platforms, and integration frameworks. 
Each compound model was trained using 5-time repeated 10-fold cross validation.
In addition, our commit classifications evaluations demonstrated $p$-value below 0.01, supporting the statistical validity of the hypothesis accuracy $>$ NIR with high confidence.\\

\noindent\textbf{\textit{Threats to Construct Validity}} consider the relationship between theory and observation, in case the measured variables do not measure the actual factors.

\begin{itemize}
    \item \underline{Manual Commit Classification}. We took the following measures to mitigate manual classification related errors:
        \begin{enumerate}
            \item Projects' issue tracking systems were used, and often provided additional information pertaining to commits.            
            \item Commits that did not lend themselves to classification due to lack of supporting information were removed from the dataset and replaced by other commits from the same repository.
            \item Both authors independently classified 10\% of the commits in the dataset used in this work. The observed agreement level was 94.5\%, and the asymptotic 95\% confidence interval for the agreement level was [90.3\%, 98.7\%] indicating that both authors agreed about the labels for the vast majority of cases.
        \end{enumerate}
    \item \underline{Source Code Change Extraction}. ChangeDistiller and the VCS mining platform we have built and used are both software programs, and as such, are not immune to bugs which could result in inaccurate or incomplete source code change extraction.
\end{itemize}

\noindent\textbf{\textit{Threats to External Validity}} consider the generalization of our findings.
\begin{itemize}
    \item \underline{Programming Language Bias}. All analyzed commits were in the Java programming language. It is possible that developers who use other programming languages, have different maintenance activity patterns which have not been explored in the scope of this work.
    \item \underline{Open Source Bias / GitHub}. The repositories studied in this paper were all popular open source projects from GitHub, selected according to the criteria described in section \ref{sec:selectingRepos}. It may be the case that developers' maintenance activity profiles are different in an open source environment when compared to other environments.
    \item \underline{Popularity Bias}. We intentionally selected the popular, data rich repositories. This could limit our results to developers and repositories of high popularity, and potentially skew the perspective on characteristics found only in less popular repositories and their developers. 
    \item \underline{Limited Information Bias}. The entire dataset, both the training and the test datasets, contained only those commits that we were able to manually classify. At the stage of VCS inspection it can be essentially impossible to actually ascertain the maintenance categories of commits that do not provide enough information traces (comment, ticket id, etc.). The true maintenance category for such commits may only be known to the developers who made them, and even they may no longer recall it soon after they have moved on to their next task.
    \item \underline{Mixed Commits}. Recent studies \cite{nguyen2013filtering,kirinuki2014hey} report that commits may involve more than one type of maintenance activity, e.g. a commit that both fixes a bug, and adds a new feature. 
    Our classification method does not currently account for such cases, but this is definitely an interesting direction to be  considered for future work (see section \ref{futureWork}).
\end{itemize}

\section{Conclusions and Future Work}
\label{futureWork}

We suggested a novel method for classifying commits into maintenance activities and used it to devise (and evaluate) a number of models that utilize commit message word frequency analysis and source code change extraction for the purpose of cross-project commit classification into maintenance activities. These models were then evaluated and compared using the accuracy and Kappa metrics with different underlying classification algorithms.
Our champion model showed a promising accuracy of 76\% and Kappa of 63\% when applied on the test dataset which consisted of 172 commits originating from various projects. These results show an improvement of over 20 percentage points, and a relative boost of over 40\% when compared to previous results (see table \ref{currentResults}), see also table \ref{tab:veryNaiveClassification} vs. table \ref{tab:bestClassification} which depict the commonly used classifier and our champion classifier, respectively.
Our work is based on studying 11 popular open source projects from various professional domains, from which we manually classified 1151 commits, $\sim$100 from each of the studied projects. The suggested models were trained using repeated cross validation on 85\% of the dataset, and the remaining 15\% of the dataset were used as a test set.

We conclude that the answer to RQ 1. is that source code changes can indeed be successfully used to devise high quality models for commit classification into maintenance activities. The answer to RQ 2. is that models that utilize source code changes are capable of outperforming the reported accuracy of word frequency based models (\cite{hindle2009automatic, amor2006discriminating}) from $\sim$60\% to $\sim$75\%, even when classifying cross-project commits.

In addition, we make the following observations based on our study:
\begin{itemize}
    \item Using text cleaning and normalization, our word frequency based models were able to achieve an accuracy of 68-69\% with Kappa of 51-53\% for cross-project commits classification  (see table \ref{tab:model_combos}).
    \item Compound models employing both (commit message) word frequency analysis and source code change types for the task of cross-project commit classification were able to achieve up to 73\% accuracy with Kappa 59\% during the training stage, and up to 76\% accuracy with Kappa of 63\% (considered ''Good`` \cite{altman1990practical}) for the test dataset.
    \item The RF algorithm outperformed the GBM and J48 in classifying cross-project commits (see table \ref{tab:testsetResults} and \ref{tab:bestClassification}).
\end{itemize}

Having an accurate classification model and the ability to apply it at scale, it may be possible to automatically classify an unprecedentedly large number of projects and commit activities. This could facilitate the revisiting of the subject of maintenance activity distribution in software projects \cite{schach2003determining, lientz1978characteristics}. 

It could also be beneficial to explore whether mixed commits (\cite{nguyen2013filtering,kirinuki2014hey}) could be automatically and accurately classified into hybrid categories, e.g. corrective$+$perfective to indicate that a given commit improves code's structure in addition to fixing a bug.

We believe that employing source code changes to help answer these questions may lead to useful insights for both practitioners and the research community.

    



\bibliographystyle{IEEEtran}
\bibliography{bibliography}

\clearpage

\end{document}